\documentclass{article}

\usepackage{arxiv}

\usepackage[T1]{fontenc}    
\usepackage{hyperref}       
\usepackage{url}            
\usepackage{booktabs}       
\usepackage{amsfonts}       
\usepackage{nicefrac}       
\usepackage{microtype}      

\usepackage{graphicx}
\usepackage{amsmath}

\title{Optimization-based approach to the calibration of mesoscale mechanical models for carbon nanotube systems}

\author{
  Vitaly Petrov\\
  Center for Design, Manufacturing and Materials\\
  Skolkovo Institute of Science and Technology,\\
  Moscow, Russia \\
  \texttt{vitaly.petrov@skoltech.ru} \\
   \And
 Igor Ostanin \\
  Center for Computational and Data-Intensive Science and Engineering\\
  Skolkovo Institute of Science and Technology\\
  Moscow, Russia \\
  \texttt{i.ostanin@skoltech.ru}
  \And
  Petr Zhilyaev\\
  Center for Design, Manufacturing and Materials\\
  Skolkovo Institute of Science and Technology,\\
  Moscow, Russia \\
  \texttt{p.zhilyaev@skoltech.ru}
}

\begin{document}
\maketitle

\begin{abstract}
A new mesoscale mechanical model, describing elastic interactions in carbon nanotubes (CNT) and other nanofilaments, is proposed. Functional form of the developed model is based on enhanced vector model (EVM) that describes basic types of bond deformations: tension, torsion, bending and shear. Calibration of bond stiffnesses is performed by adjusting EVM parameters to reproduce both CNT's deformation energies and shape observed in a full-atomistic simulation. The parameters obtained are compared with the ones obtained from Euler-Bernoulli beam theory considerations. It is found that after certain critical length of a tested CNT specimen its stiffness parameters become length-independent and can be used in mesoscale simulations of CNTs of arbitrary length.
\end{abstract}


\section{Introduction}
Since their discovery~\cite{iijima1991helical}, carbon nanotubes (CNT) became a core component for a great variety of technologies, ranging from reinforced composites~\cite{esawi2007carbon} to wearable electronics devices~\cite{gilshteyn2018flexible}. \footnote{Although in the following we limit our discussion to CNTs, our results are directly applicable to other types of nanofilaments, such as boron-nitride nanotubes and zinc oxide nanotubes that are actively studied in a context of numerous possible applications. } Further technological development urges the design of predictive models that could simulate the behaviour of CNT-based materials. Although \textit{ab-initio} and atomistic modelling give deep insight into underlying physics and mechanics of CNTs, many systems of scientific and engineering significance (\textit{e.g.} CNT films, ropes, mats and forests) remain computationally prohibitive. In order to simulate such systems, one has to use mesoscale/coarse-grained (CG) simulation approaches.

In order to provide an efficient and accurate description of the mechanics and physics of CNT systems, a number of different mesoscale models were developed. One of the first was bead and spring (BS) model \cite{buehler2006mesoscale}, which represents a CNT as the chain of point masses (beads). Within this approach a "spring" potential is used to describe covalent interactions within a CNT, and a coarse-grained Lennard-Jones potential is employed to model dispersive interactions between neighbouring CNTs. Such model can easily be implemented in existing molecular dynamics codes, and it reasonably well reproduces the mechanical properties of a single CNT. However, this modeling approach has a list of critical shortcomings that make it inapplicable in many relavant situations. The first of them is the absence of torsional degrees of freedom that produces unrealistic behaviour of CNT assemblies under certain loadings. The second problem is the artificial corrugation of van der Waals interaction potential, emerging due to representation of a cylindrical CNT segment with a point mass. Such an artefact leads to the fact that BS model can not reproduce realistic self-assembly processes in CNT systems due to vdW adhesion, and grossly overestimates yield strength of CNT materials (see the discussion in []).   

Another CG approach for modelling CNT, mesoscopic force-field (MFF), is proposed by Zhigilei et al.~\cite{zhigilei2005mesoscopic}. In this framework, CNT dynamics also described in terms of nodal point masses. Although in the seminal work~\cite{zhigilei2005mesoscopic} torsion term is present in the Lagrangian of the conceptual CNT elasticity model, the authors later neglect it and coarse-grain CNT description to the level of a chain of point masses. In contrast to BS, further development of the MFF approach with inter-tube interactions~\cite{volkov2008mesoscopic} allowed to simulate formation of large bundles composed of approximately 50 -- 100 CNTs, which is qualitatively consistent with experiments~\cite{hobbie2010wrinkling}. 

Torsion part of CNT's potential energy is consistently taken into account in the study~\cite{ostanin2013distinct}. Proposed CG model is based on parallel-bond (PB) representation of the potential energy~\cite{potyondy2004bonded} and represents nanotube as a chain of rigid bodies with both positions and orientations. This allows to describe not only bending but also torsion of a single CNT, which is important for modelling ropes and fibers woven of individual CNTs \cite{wang2015twisting}. 

A common issue of all listed approaches which has not been addressed so far is the rigorous and accurate procedure of coarse-grained model`s calibration based on molecular level simulations.
In all the described approaches \cite{buehler2006mesoscale, zhigilei2005mesoscopic, ostanin2013distinct}, the model for nanotube stiffness utilises linear elastic beam theory. Clearly, such "top-down" approach involves an unsupported assumption of the beam`s theory validity on the microscopic level and can lead to significant model's inaccuracy. Although in work~\cite{zhigilei2005mesoscopic} deformation energies for the calibration were calculated on atomistic level, comparison between the shape of mesoscale and full-atomic CNTs were not performed.  

In this work we present a new approach to the construction of a mesoscale mechanical model for CNT. It is based on the concurrent energy minimization in the coarse and fine models with subsequent  matching potential energies of the system. This approach is employed for calibration of recently developed enhanced vector model (EVM)~\cite{kuzkin2012vector}. Unlike previously used parallel bond model, it provides precise energy conservation in zero-damping simulation, and can be easily generalised to capture nonlinear effects and bond fracture. Our approach completely relies on atomistic-level computations, none of the results from linear elastic beam theory are used in the calibration procedure.

Our results allowed to improve calibration accuracy for the recently developed mesoscale distinct element method (MDEM) \cite{ostanin2018toward} for large-scale dynamic modelling of carbon nanotube assemblies. The proposed framework can also be used in more general context for calibration and verification of CG models for different types of nanostructures.

\begin{figure}[h!]
    \centering
    \includegraphics[scale=0.5]{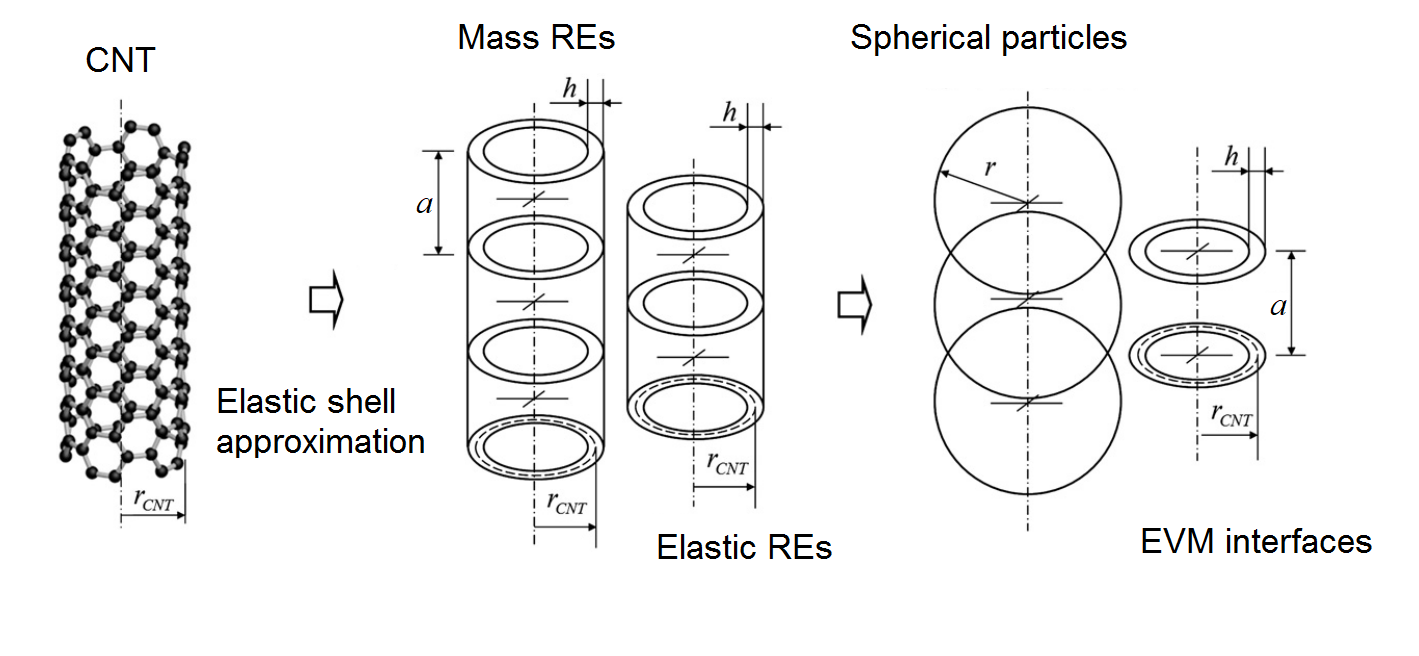}
    \caption{Illustration of CNT coarse-graining procedure}
    \label{figures::CNT_coarse_graining}
\end{figure}

\section{Methods}

\subsection{Mesoscale model}

Within our mesoscale model, a CNT is partitioned into a chain of identical hollow cylindrical segments, representing inertial and elastic properties of the corresponding groups of carbon atoms (fig. \ref{figures::CNT_coarse_graining}). Each segment has a length of $a$ equal to two CNT radii: $a = 2r_{CNT}$.  

\begin{figure}[!ht]
    \centering
    \includegraphics[scale = 0.35]{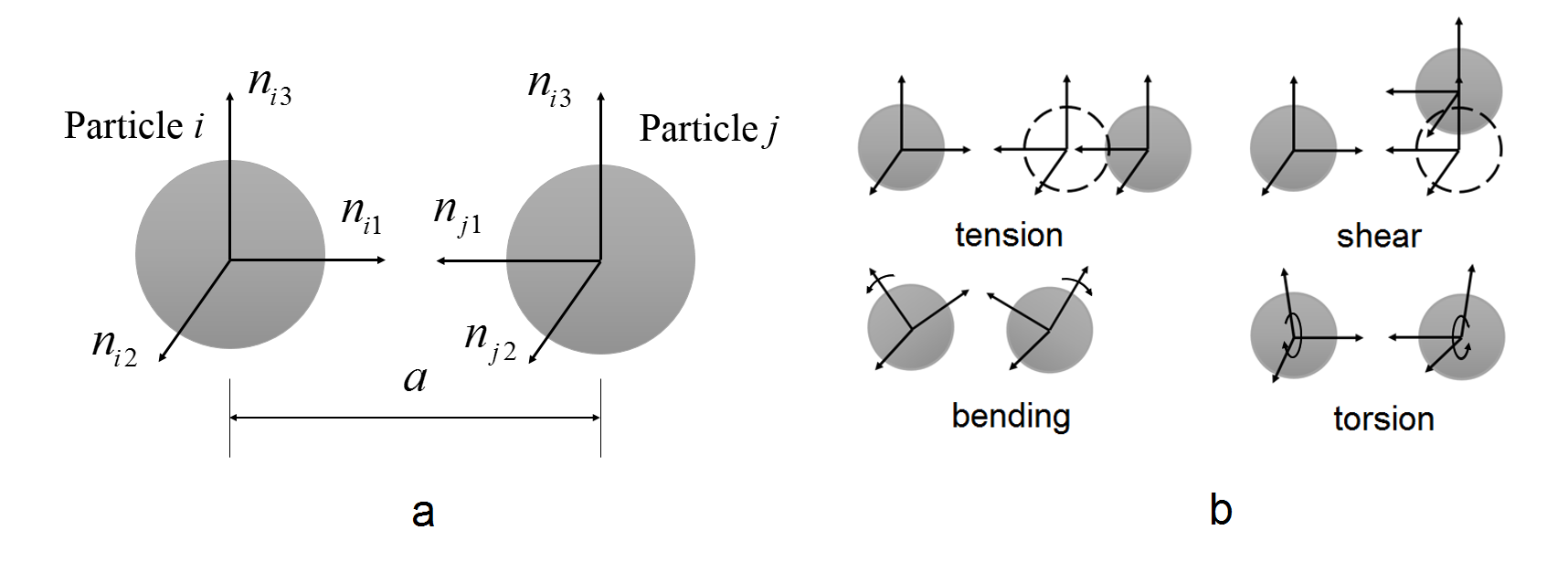}
    \caption{\textbf{a)} Example of two rigid particles linked with EVM bond; \textbf{b)} Four types of bond deformation. }
    \label{figures::v_bond_concept}
\end{figure}

The inertial properties of the segment are represented with the rigid spherical particle. Its radius is chosen to match the moment of inertia of the segment with respect to CNT axis. The corresponding inertial radius is given by $r=\sqrt{2.5} r_{CNT}$.  

The elastic properties of a segment are described by EVM~\cite{kuzkin2012vector, kuzkin2017enhanced}. It is based on a binding potential, describing the behaviour of an elastic material linking two rigid bodies. The formulation provides straightforward generalisation on the case of large strains and accounts for a bending-twisting geometric coupling. Consider two equal-sized spherical particles $i$ and $j$ with equilibrium separation distance $a$ and orientations described by orthogonal vectors $\textbf{n}_{ik}, \textbf{n}_{jk}$ (fig. \ref{figures::v_bond_concept} (a) , $\textbf{n}_{i1} = - \textbf{n}_{j1}$, $\textbf{n}_{ik} =  \textbf{n}_{jk}$, $k = 2, 3$ ).

The potential energy in EVM is given by:
\begin{equation}\label{equations::v_bond_model}
    U = \sum_{bonds}  \frac{B_1}{2} \left( D_{ij} - a \right)^2 + \frac{B_2}{2} \left( \textbf{n}_{j1} \textbf{n}_{i1} \right) \cdot \textbf{d}_{ij} - B_3 \textbf{n}_{i1} \cdot \textbf{n}_{j1} - \frac{B_4}{2} \left( \textbf{n}_{i2} \cdot \textbf{n}_{j2} + \textbf{n}_{i3} \cdot \textbf{n}_{j3} \right).
\end{equation}
Here $\textbf{D}_{ij}$ is a radius-vector connecting centers of bonded particles, $\textbf{d}_{ij} = \textbf{D}_{ij}/  \lvert \textbf{D}_{ij} \rvert\ $.

The structure of (\ref{equations::v_bond_model}) allow us to consider it as a combination of four elementary bond`s deformations: tension, shear, bending and torsion (fig. 2 \ref{figures::v_bond_concept}, (b)). Thereby parameters $B_1, B_2, B_3$ \text{and} $B_4$ are directly related to longitudinal, shear, bending, and torsional stiffnesses of a bond respectively.

\subsection{Atomistic model}
Atomistic simulation is used to define the difference ( $\Delta U_{atomistic}$) between energies in unstained and deformed states. This value is then utilized in calibration of mesoscale model parameters as described below.

Considered atomistic configurations of CNT contain from $10^4$ to $10^5$ carbon atoms. Interactions between neighbouring atoms are described by AIREBO (Adaptive Intermolecular Reactive Empirical Bond Order) \cite{stuart2000reactive} potential. HFTN (Hessian Free Truncated Newton)  optimization algorithm \cite{nash2000survey} is used for structure relaxation. Non-periodic boundary conditions are imposed.

In order to provide one-to-one correspondence between atomistic and coarse-grained models, we unite carbon atoms into groups corresponding to the segments in a mesoscale model. During our simulations, we track the averaged group positions, that are then used in comparison of the shapes obtained from atomistic and mesoscale models.

\subsubsection{Calibration}
EVM potential (\ref{equations::v_bond_model}) allows to consider an arbitrary deformation as a combination of four elementary modes -- tension, shearing, bending and torsion. Hence, EVM calibration on the results of atomistic simulation can be carried out by four concurrent elementary tests in CG and full-atomistic models and direct matching of deformation energies in all cases.

Such a strategy is described with the following optimization problem:
\begin{equation}\label{equations::optimization_problem}
        Q(B_1,.., B_4) = \left( \Delta U_{\text{atomistic}} - \Delta U_{\text{CG}} \left( B_1,.., B_4 \right) \right)^2 \longrightarrow min.
\end{equation}
Here $Q(B_1,.., B_4)$ is an error function, $\Delta U = U_{deformed} - U_{initial}$ -- energy of deformation in atomistic and CG models.

Calibration procedure can be described as a three-stage process:

(I) On the first stage, the atomistic simulation is used to compute a set of deformation energies $\Delta U_{\text{atomistic}}$ and positions for different types of loading.

(II) The next step is a static energy minimization of CG CNT configuration. At this stage the parameters $B_1, ..., B_4$ are fixed and only coarse-grained degrees of freedom (positions and rotations) are varied to obtain a minimum of potential energy. As a result one can obtain a certain CNT configuration which is identical to the solution of static equilibrium equations for CNT. Minimization is performed by  basin hopping optimization technique \cite{olson::2012::basinhopping}. 

(III) The last stage is adjusting $B_1, ..., B_4$ coefficients while solving problem (\ref{equations::optimization_problem}) utilizing differential evolution \cite{Storn::1997::differential_evolution} algorithm. 

As the result of this process, we obtain the CG potential that reproduces both deformation shapes and energies observed in full atomistic simulation.

\section{Results}

In this section we demonstrate the application of the proposed calibration procedure to single-walled CNTs of various lengths and chiralities. We first obtain the initial guess for CG model parameters from linear Euler-Bernoulli beam theory. Then we follow the calibration algorithm described above. The coefficients $B_1$ (tension) and $B_4$ (torsion) of EVM model are obtained from decoupled deformation tests, while $B_2$ (shear) and $B_3$ (bending) stiffnesses are evaluated from coupled deformations. The pipeline is demonstrated in details for (10,10) CNT, the results for different types of CNTs are briefly presented in the last subsection.  

\subsection{Initial guess}

In the first approximation the CG parameters are estimated using Euler-Bernoulli beam theory~\cite{Kuzkin::2012::Vmodel_initial}:

\begin{equation}\label{equations::rod_calibration}
        \begin{split}
            & B_1 = \frac{ES}{a},\\[5pt]
            & B_2 = \frac{12EJ}{a},\\[5pt]
            & B_3 = \frac{2EJ}{a} + \frac{G J_p}{2a},\\[5pt]
            & B_4 = \frac{G J_p}{a}.
        \end{split}
\end{equation}

Here $a$, $S$, $J$ and $J_{p}$ are equilibrium distance between segments, interface cross section area, moment of inertia and polar moment of inertia respectively:

\begin{equation}\label{equations::area_and_moments_definitions}
    \begin{split}
        & S = 2 \pi h r_{\text{CNT}},\\
        & J = \pi h r_{\text{CNT}} \left( r^2_{\text{CNT}} + \frac{h^2}{4} \right),\\
        & J_{p} = 2 J,
    \end{split}
\end{equation}.

Consider the case of (10, 10) CNT which is divided into segments containing approximately 220 carbon atoms. The comparison between segments in full-atomic and mesoscale models are carried out with respect to it`s length, hence there is no uniform distribution of atoms into segments. Mass and geometry properties of a segment are presented in table \ref{table::elastic_params}; Young's modulus $E = 1029 \text{ GPa}$ and shear modulus $G = 459 \text{ GPa}$ are taken from \textit{ab-initio} calculations~\cite{Dumitrica::2008::atomistic_cnt}.

\begin{table}[h!]
    \centering
    \begin{tabular}{|c|c|c|c|c|c|c|c|c|c|}
    \hline
    $m, \text{amu}$ & $I, \text{amu} \cdot \text{\AA}^2$ & $a, \text{\AA}$ & $h, \text{\AA}$ & $S, \text{\AA}^2$ & $J, \text{\AA}^4$ & $J_p, \text{\AA}^4$\\
    \hline
    2649 & $1.22 \cdot 10^5$ & 13.56 & 3.35 & 142.7 & 3480 & 6960\\
    \hline
    \end{tabular}\\[10pt]
    \caption{Mass and geometry parameters of segments for (10, 10) CNTs.}
    \label{table::elastic_params}
\end{table}

Utilizing this data, one can estimate $B_1 = 67.6 \text{ eV} / {\text{\AA}^2}$, $B_2 = 19800 \text{ eV}$, $B_3 = 4030 \text{ eV}$, $B_4 = 1400 \text{ eV}$. Obtained values are used as an initial guess in our optimization-based calibration procedure.

\subsection{Tensile stiffness calibration}

Consider pure stretching of CNT along the axial vector $\textbf{D}_{ij}$,  which is equivalent to tension of all bonds between segments within considered CNT. In this case the potential (\ref{equations::v_bond_model}) can be simplified to:

\begin{equation}
\label{equations::v_bond_model_stretching}
    U = \sum_{bonds}  \frac{B_1}{2} \left( D_{ij} - a \right)^2.
\end{equation}

In order to provide a uniform stretching of all bonds in CNT we displace the last segment/group of atoms in the chain to the distance of $\Delta \ll a$ at both CG and atomistic models. Then positions of the first and last segment`s/groups of atoms are fixed so the other segments/atoms change their location during optimization procedure in order to minimize total potential energy. As a result of the energy minimisation all bonds of the CNT are become equally stretched in both models (fig. \ref{figures::stretching_atomistic_and_mesoscale}):

\begin{figure}[!ht]
    \centering
    \includegraphics[scale = 0.475]{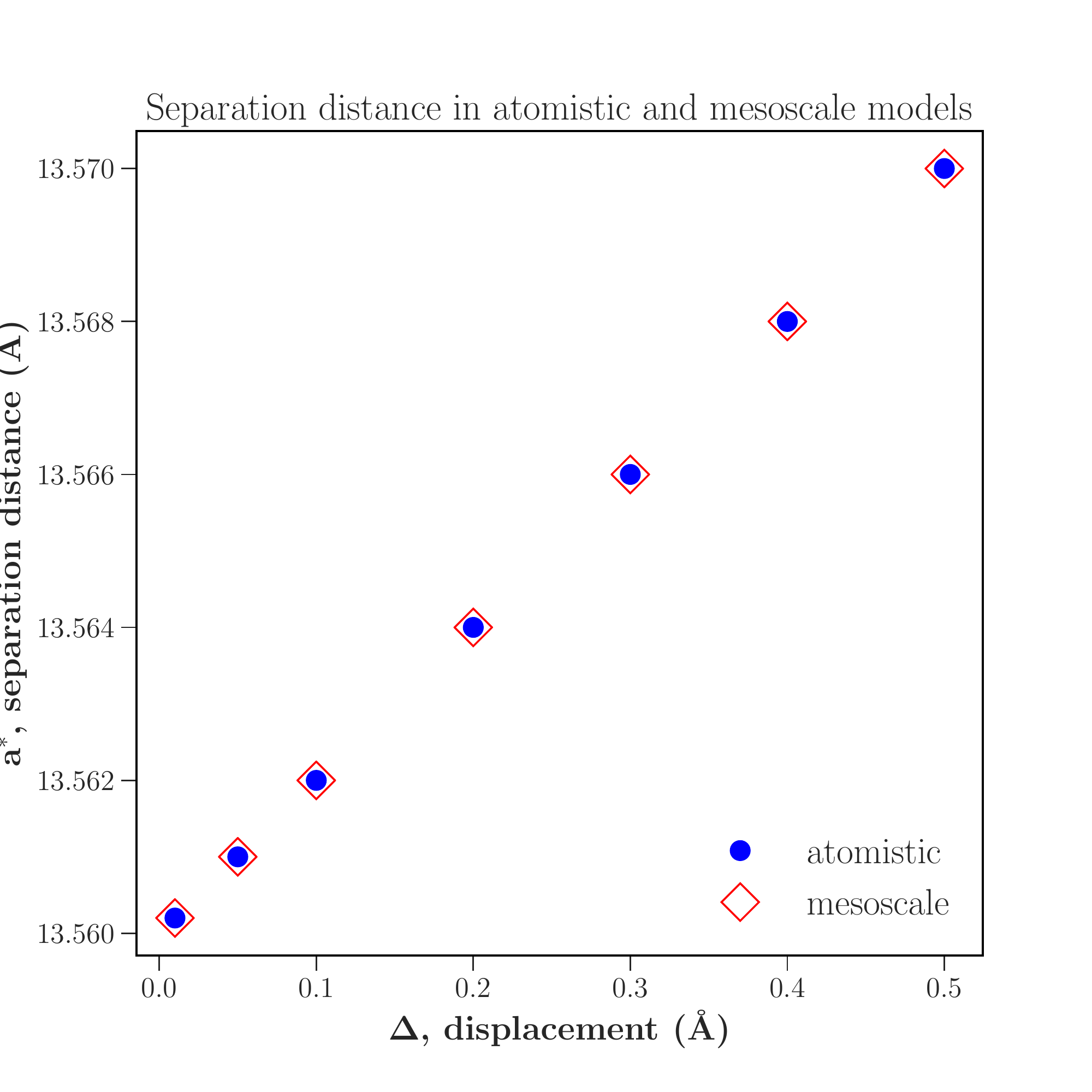}
    \caption{Pair separation distance for (10, 10) CNT as a function of displacement for last segment/group of atoms in chain. }
    \label{figures::stretching_atomistic_and_mesoscale}
\end{figure}

The optimization problem (\ref{equations::optimization_problem}) is solved with simplified potential (\ref{equations::v_bond_model_stretching}) and $B_1$ is found to be equal to $60.1 \text{ eV} / {\text{\AA}^2}$. For long enough CNTs, the value of $B_1$ does not depend on the CNT length and magnitude of the displacement (see the discussion below). 

\subsection{Torsional stiffness calibration}

For the case of pure axial torsion the potential (\ref{equations::v_bond_model}) can be simplified to:
\begin{equation}
    \label{equations::v_bond_model_twisting}
    U = \sum_{bonds} -\frac{B_4}{2} \left( \textbf{n}_{i2} \cdot \textbf{n}_{j2} + \textbf{n}_{i3} \cdot \textbf{n}_{j3} \right)
\end{equation}

Torsion test is performed by rotating of the last segment/group of atoms on the angle of $\varphi$ along the nanotube axis. Orientation of the first and last segments / group of atoms are then fixed and configuration optimization is performed. As a result, one can obtain a linear distribution of twisting angle along the CNT axis (fig. \ref{figures::torsion_distribution}):

\begin{figure}[!ht]
    \centering
    \includegraphics[scale = 0.475]{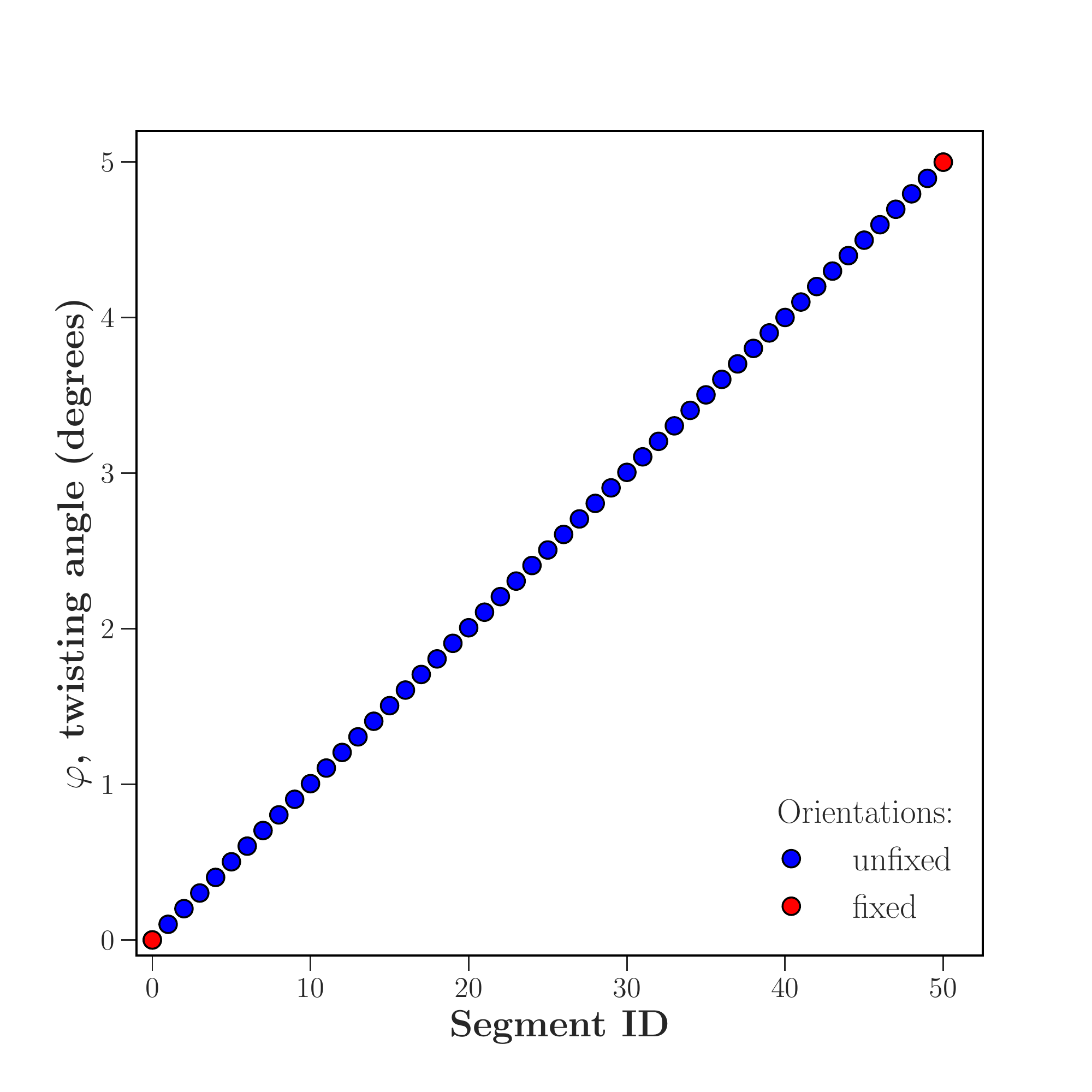}
    \caption{Segment`s twisting angle distribution within CNT. The linear nature of the obtained dependency indicates a validity of our optimization technique}
    \label{figures::torsion_distribution}
\end{figure}

Note that the results of structure optimization for tensile and torsion tests are obvious and follow directly from the balance of force and moment in a chain of interacting segments. We provide them as benchmarks for our calibration technique. The resulting configurations are in perfect agreement with analytical solutions, which verifies the validity of our approach.

The optimization problem (\ref{equations::optimization_problem}) is solved with simplified potential (\ref{equations::v_bond_model_twisting}), as a result we found $B_4$ coefficient to be equal to 1107 eV. Similarly to the stretching case, the obtained value of twisting stiffness $B_4$ does not depend both on deformation value and CNT length (for long enough CNTs), as discussed below.

\subsection{Shearing and bending cases}

So far we considered only decoupled deformation cases that admitted single-parameter optimization. However, this approach is inapplicable for bending/shear deformation. In this case, the potential (\ref{equations::v_bond_model}) will take a form of:

\begin{equation}
    \label{equations::v_bond_model_shear_and_bending}
    U = \sum_{bonds} \left( \frac{B_2}{2} \left( \textbf{n}_{j1} - \textbf{n}_{i1} \right) \cdot \textbf{d}_{ij} - B_3 \textbf{n}_{i1} \cdot \textbf{n}_{j1} \right)
\end{equation}

Potential (\ref{equations::v_bond_model_shear_and_bending}) depends both on shear stiffness $B_2$ and bending stiffness $B_3$ which can not be decoupled.

For this case the optimization problem (\ref{equations::optimization_problem}) is slightly modified in order to minimize error in potential energy at two bending/shear tests by fitting two stiffness parameters:

\begin{equation}
    \label{equations::optimization_problem_modified}
    Q_{\text{1}}^2 \left( B_2, B_3 \right) + Q_{\text{2}}^2 \left( B_2, B_3 \right) \longrightarrow min.
\end{equation}
Here $Q_{\text{1}}$, $Q_{\text{2}}$ are error functions for two independent tests - three-point and similar two-point coupled bending/shearing tests.
 
In the first case the positions and orientations of two edge segments/group of atoms of a CNT specimens are fixed. Then the N /2 and N / 2 + 1 (two segments are taken in order to conserve symmetry) segments/groups of atoms are displaced in transversal direction to the value of $\Delta y \ll a$. The optimization process converges to an equilibrium CNT profile (fig. \ref{figures::bending_and_shearing_profiles}, a):

\begin{figure}[!ht]
    \begin{minipage}{0.4\linewidth}
        \centering
        \includegraphics[scale = 0.4]{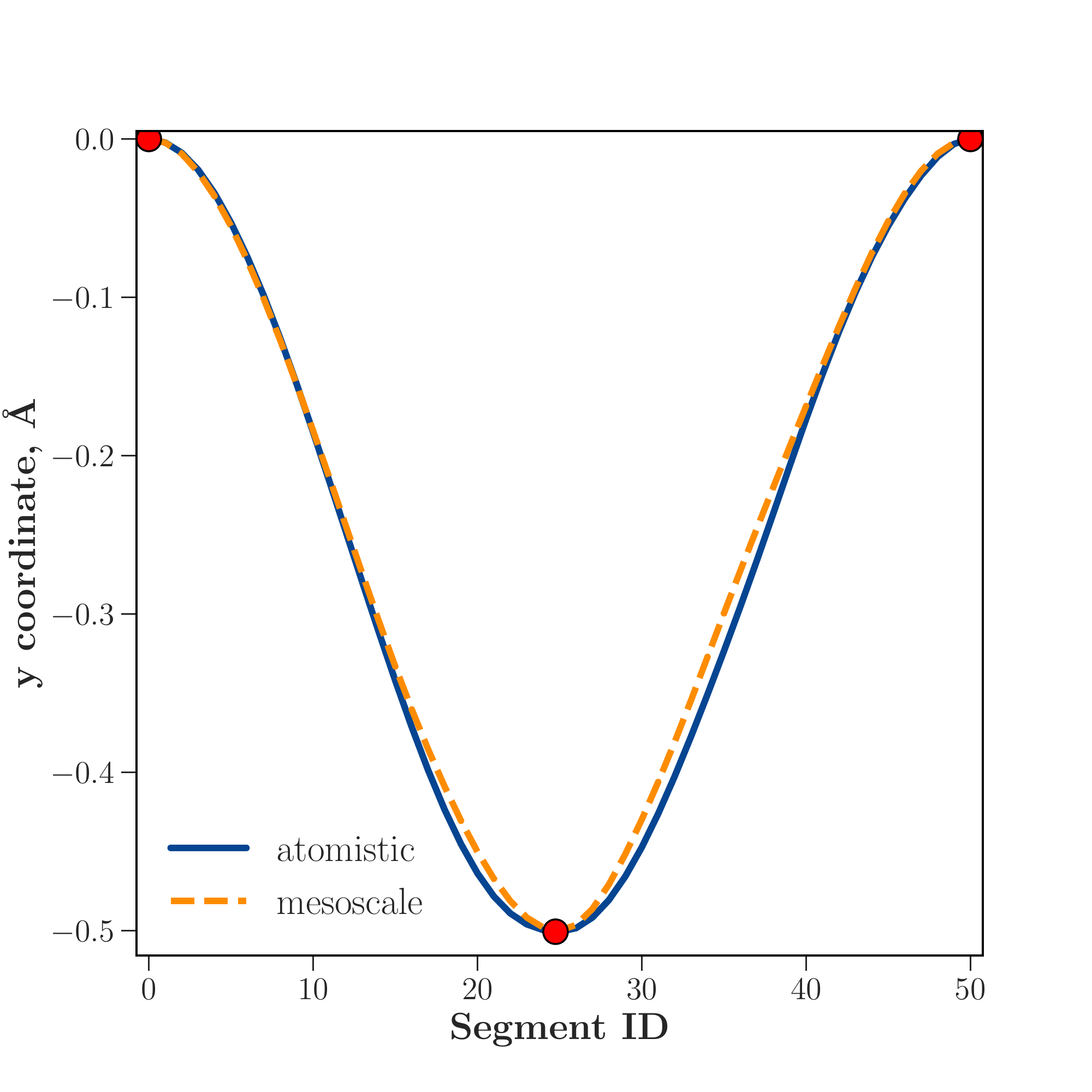} a)
    \end{minipage}
    \begin{minipage}{0.4\linewidth}
        \centering
        \includegraphics[scale = 0.4]{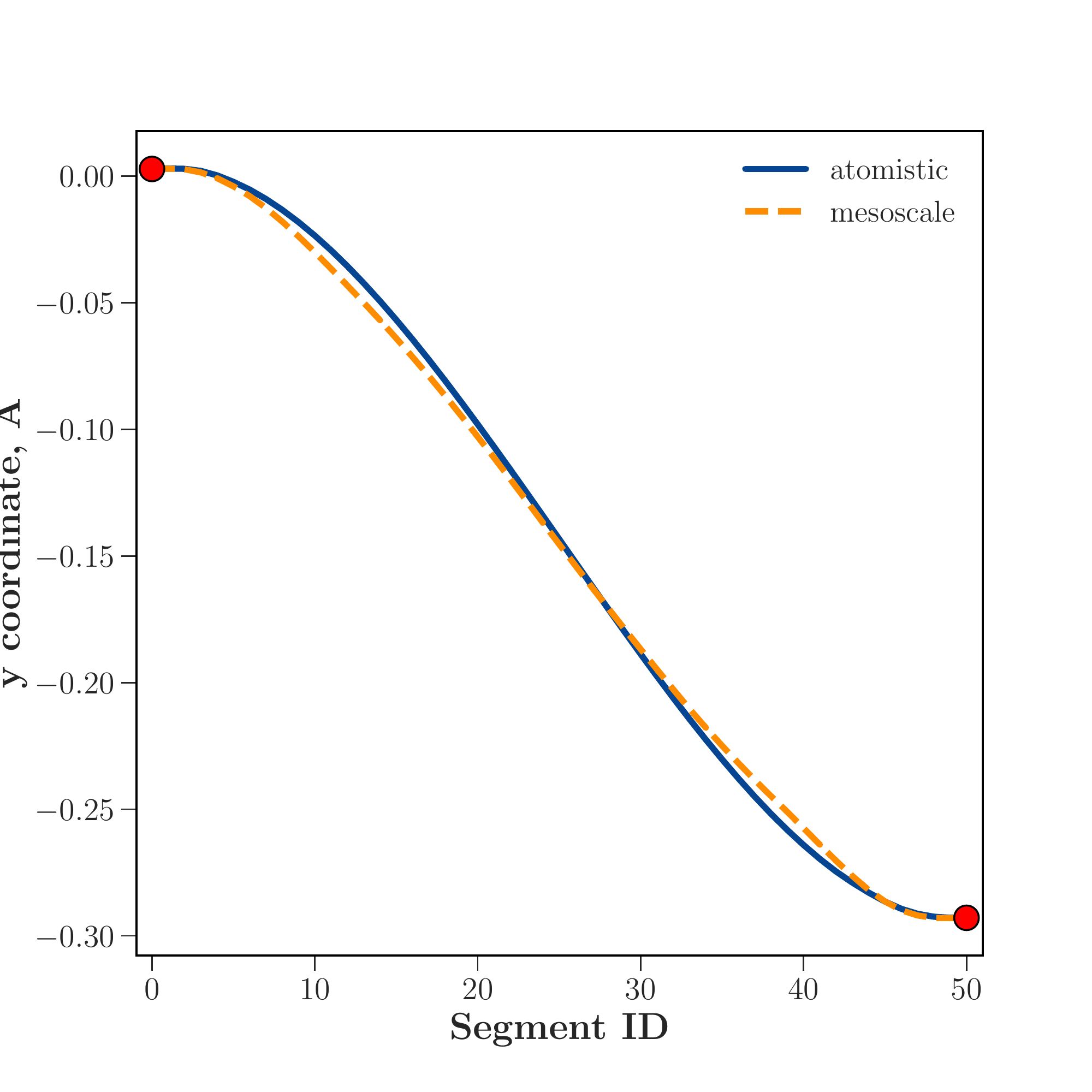} b)
    \end{minipage}
    \caption{Bended and sheared CNT profiles in full-atomic and CG models. Red dots -- segments with fixed positions and orientations.}
    \label{figures::bending_and_shearing_profiles}
\end{figure}

In the second case, the similar test is performed in two point manner: the first segment/group of atoms is fixed, and the last one is displaced on $\Delta y \ll a$ (fig. \ref{figures::bending_and_shearing_profiles}, b)

While solving (\ref{equations::optimization_problem_modified}) with the potential (\ref{equations::v_bond_model_shear_and_bending}), one can found for $B_2$ and $B_3$ to be equal to 17100 eV and 3610 eV respectively. 

Our concurrent optimization approach leads to the set of stiffness values $B_1,..B_4$ that are noticeably different from the predictions of beam theory:

\begin{table}[!ht]
    \centering
    \begin{tabular}{|c|c|c|c|c|c|c|c|c|c|}
    \hline
    & Optimization-based & Euler-Bernoulli\\
    \hline
    $B_1, \text{ eV} / {\text{\AA}^2}$ & 60.1 & 67.6\\
    \hline
    $B_2, \text{ eV}$ & 17100 & 19800\\
    \hline
    $B_3, \text{ eV}$  & 3610 & 4030\\
    \hline
    $B_4, \text{ eV}$  & 1107 & 1400\\
    \hline
    \end{tabular}\\[10pt]
    \caption{Elastic potential (\ref{equations::v_bond_model}) calibration and compairison with theoretical estimation.}
    \label{table::calibration_results}
\end{table}

\subsection{Other types of CNTs}

We performed similar calibrations for two other types of CNTs:
(10, 0) "zigzag" CNT`s and (15,15) "armchair" CNTs. 

Table  summarizes EVM parameterizations for three different CNT types and segment sizes. 
\begin{table}[!ht]
    \centering
    \begin{tabular}{|c|c|c|c|c|c|c|c|c|c|}
    \hline
    Segment length, \AA & 6.78 & 13.56 & 27.12\\
    \hline
    Coefficient $B_1, \text{ eV} / {\text{\AA}^2}$ & 115.9 & 60.1 & 30.9\\
    \hline
    Coefficient $B_2, \text{ eV}$ & 35500 & 17100 & 9300\\
    \hline
    Coefficient $B_3, \text{ eV}$ & 10660 & 3610 & 1850\\
    \hline
    Coefficient $B_4, \text{ eV}$ & 2228 & 1107 & 592\\
    \hline
    \end{tabular}\\[10pt]
    \caption{Dependence of elastic stiffnesses value on segment size.}
    \label{table::segment_size_dependence}
\end{table}

Tables give comparison of initial guess values with optimization results for (10,0) CNTs and (15,15) CNTS.

\begin{table}[!ht]
    \centering
    \begin{tabular}{|c|c|c|c|c|c|c|c|c|c|}
    \hline
    & Optimization-based & Euler-Bernoulli\\
    \hline
    $B_1, \text{ eV} / {\text{\AA}^2}$ & 41.1 & 28.0\\
    \hline
    $B_2, \text{ eV}$ & 5850 & 2700\\
    \hline
    $B_3, \text{ eV}$  & 1340 & 550\\
    \hline
    $B_4, \text{ eV}$  & 230 & 100\\
    \hline
    \end{tabular}\\[10pt]
    \caption{Elastic potential (\ref{equations::v_bond_model}) calibration and compairison with theoretical estimation for (10, 0) CNT`s.}
    \label{table::calibration_results_zigzag}
\end{table}

\begin{table}[!ht]
    \centering
    \begin{tabular}{|c|c|c|c|c|c|c|c|c|c|}
    \hline
    & Optimization-based & Euler-Bernoulli\\
    \hline
    $B_1, \text{ eV} / {\text{\AA}^2}$ & 85.1 & 98.2\\
    \hline
    $B_2, \text{ eV}$ & 78100 & 106200\\
    \hline
    $B_3, \text{ eV}$  & 17500 & 21650\\
    \hline
    $B_4, \text{ eV}$  & 3411.1 & 3950.6\\
    \hline
    \end{tabular}\\[10pt]
    \caption{Elastic potential (\ref{equations::v_bond_model}) calibration and compairison with theoretical estimation for (15, 15) CNT`s.}
    \label{table::calibration_results_15_15}
\end{table}

\subsection{Size dependence of stiffnesses}

Our tests indicate that short CNTs feature significant size dependence of stiffness (fig. \ref{figures::coeffs_deviation}). However, for long enough CNTs (approximately 50 nm for (10,10) CNTs), this size dependence vanishes, and stiffness calibrations do not depend on the specimen's size. 

\begin{figure}[!ht]
    \begin{minipage}[h!]{0.5\linewidth}
        \center{\includegraphics[width=0.75\linewidth]{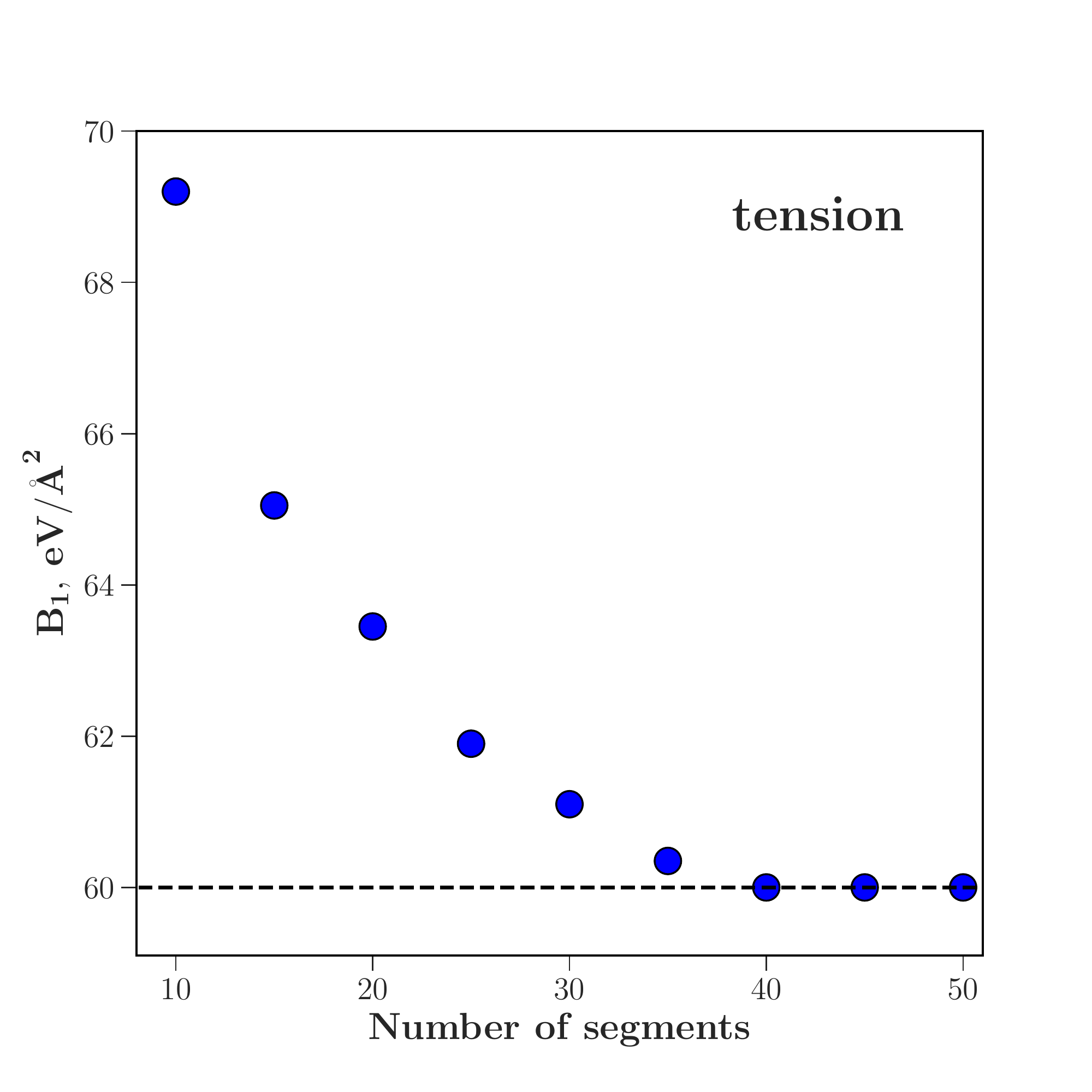}} a)
    \end{minipage}
    \hfill
    \begin{minipage}[h!]{0.5\linewidth}
        \center{\includegraphics[width=0.75\linewidth]{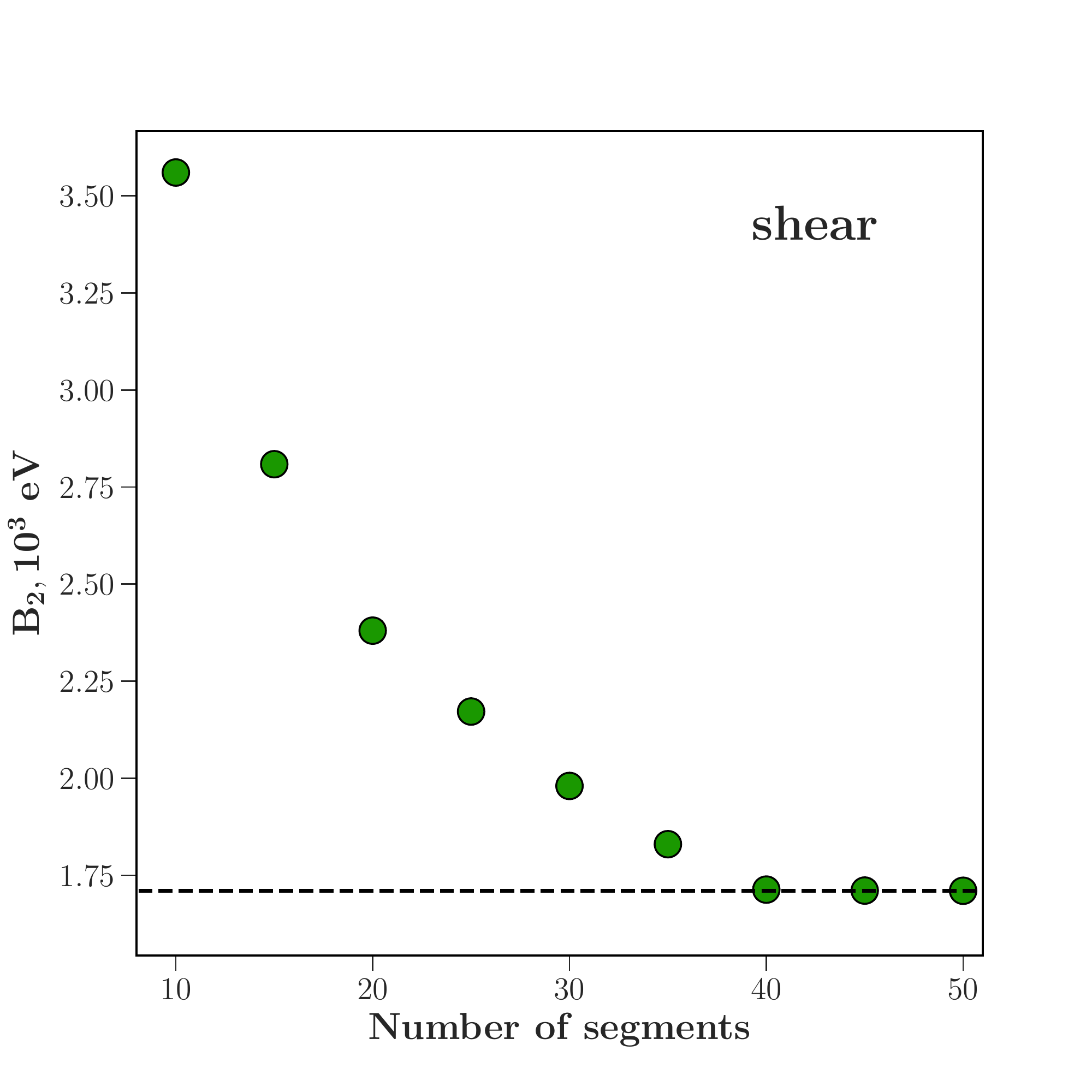}} b)
    \end{minipage}
\end{figure}

\begin{figure}[!ht]
    \begin{minipage}[!ht]{0.5\linewidth}
        \center{\includegraphics[width=0.75\linewidth]{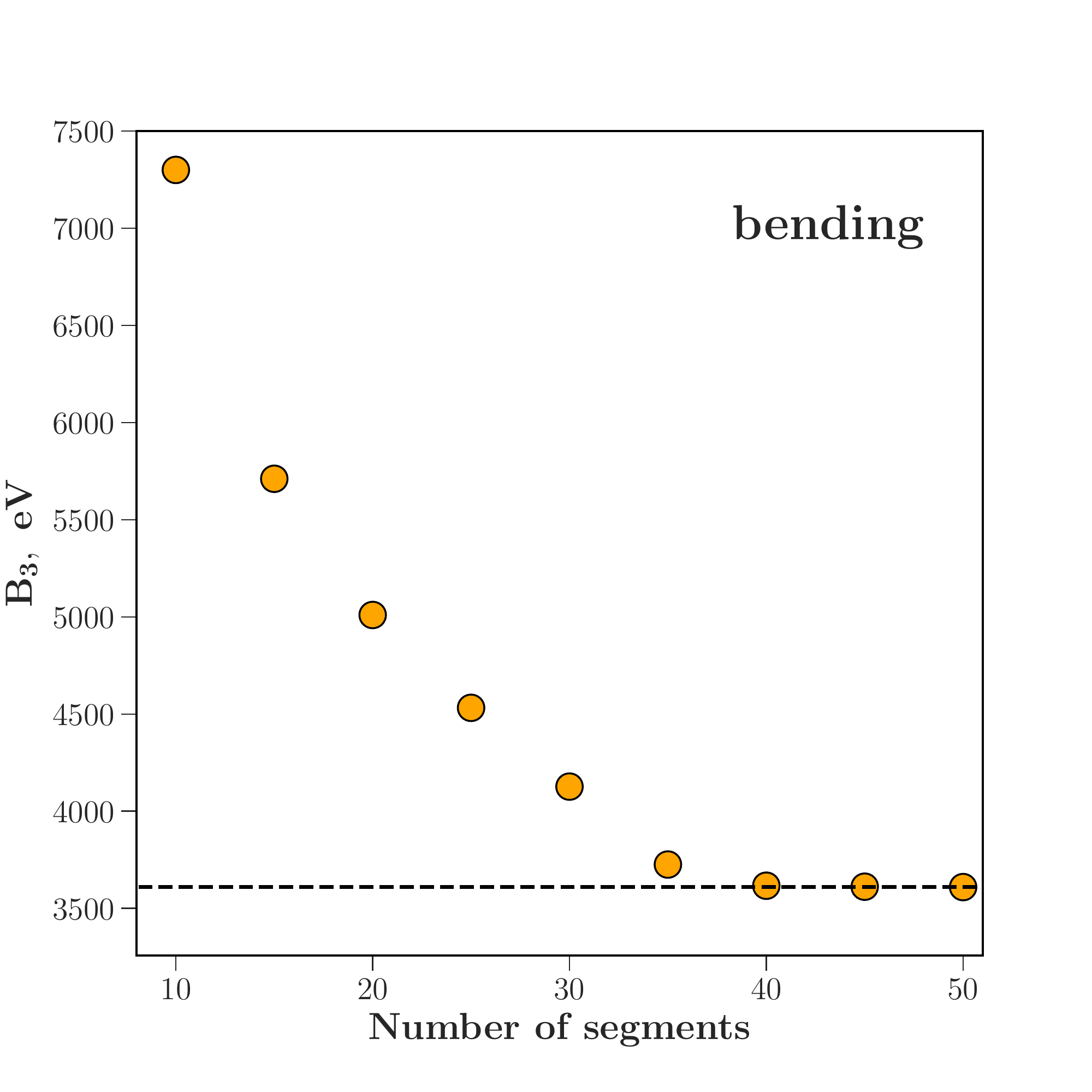}} c)
    \end{minipage}
    \hfill
    \begin{minipage}[!ht]{0.5\linewidth}
        \center{\includegraphics[width=0.75\linewidth]{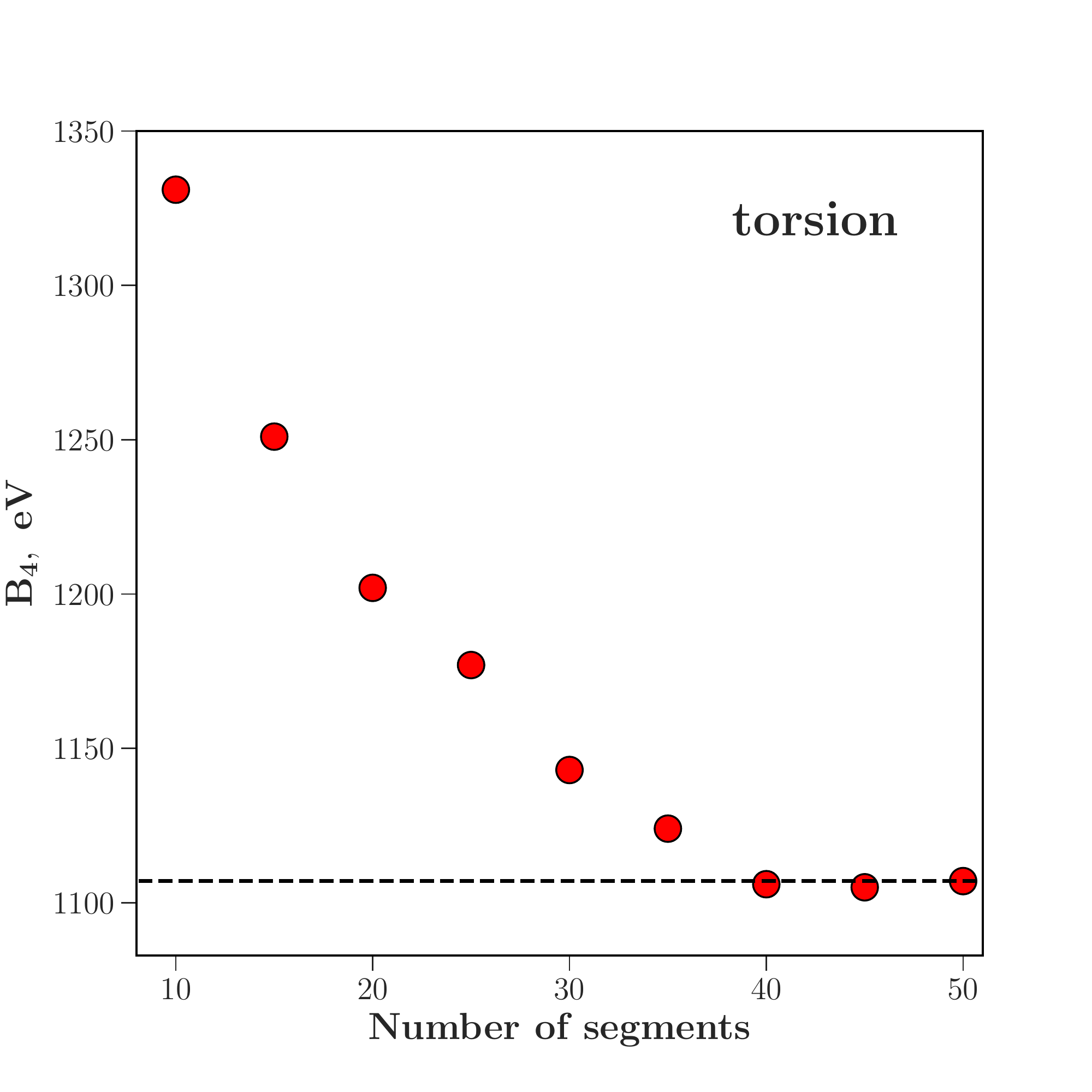}} d)
    \end{minipage}
    \caption{Deviations from beam`s theory in case of short nanotubes for a) stretching, b) shear, c) bending and d) twisting stiffnesses.}
    \label{figures::coeffs_deviation}
\end{figure}
\newpage

\section{Conclusions}

In this work we have described a systematic "bottom-up" approach to calibration of mesoscale mechanical models for elastic fibrils based on optimization technique. We illustrated our approach by calibrating tension, shear, bending and torsional stiffnesses for CG models of few types of CNTs. It was shown that mesoscopic EVM potential, which was used for elastic interactions, does reproduce fibril's deformation. Our approach allows to refine the calibrations used in mesoscale modeling of nanofilaments, and can be straightforwardly generalized onto different types of mesoscale structures (\textit{e.g} ZnO nanobelts).

\section{Acknowledgements}

The authors gratefully acknowledge the financial support from Russian Science Foundation under grant 17-73-10442.

\bibliographystyle{unsrt}  


\bibliographystyle{natbib}
\bibliography{references} 

\end{document}